# Observational support for the Gurzadyan-Kocharyan relation in clusters of galaxies

Pierluigi Monaco*

[1] Scuola Internazionale Superiore di Studi Avanzati (SISSA), via Beirut 4, 34013 Trieste, Italy
[2] Dipartimento di Astronomia, Università degli Studi di Trieste, via Tiepolo 11, 34131 Trieste, Italy

**Abstract.** We show that observational data for four Abell clusters of galaxies support the Gurzadyan-Kocharyan relation between the Hausdorff dimension and the dynamical properties of a galaxy system. The Hausdorff dimension is calculated using the two-point correlation function, while the dynamical parameters are estimated using available data and reasonable assumptions on the mass function of galaxies. This result can have essential consequences in the understanding of the dynamical mechanisms that determine the fractal distribution of galaxies.

## 1. Introduction

A fundamental feature of our observable Universe is the evidence for the fractal nature of the distribution of galaxies. It is well known that the two-point correlation function $\xi(r)$ is well-described, on small scales, by a power-law (see, e.g., Peebles 1993):

$$\xi(r) \propto r^{-\gamma}, \qquad \gamma \simeq 1.8, \qquad (1)$$

where the exponent $\gamma$ is related to the Hausdorff (fractal) dimension by:

$$d_H = 3 - \gamma \simeq 1.2. \qquad (2)$$

This behaviour has been found to hold up to a few times the correlation length $r_0 \simeq 5\ h^{-1}$ Mpc, that is the scale at which $\xi = 1$ (here $H_0 = 100 h^{-1}$ km s$^{-1}$ Mpc$^{-1}$). At the same time, clusters of galaxies show a similar fractal distribution, with the same Hausdorff dimension (i.e. the same $\gamma$), but with a correlation length four or five times greater, about $20/25\ h^{-1}$ Mpc (see, e.g., Bahcall 1988). This difference can be explained by means of a "biased" galaxy formation scenario (see Kaiser 1984; Bardeen et al. 1986). Alternatively the definition of the correlation length has been criticized by Coleman & Pietronero (1992), who have argued that the galaxy distribution is fractal up to scales at least as large as $30\ h^{-1}$ Mpc.

From the theoretical point of view, the meaning of the Hausdorff dimension of a galaxy distribution is debated. In the linear regime, when the primordial perturbations have grown by a small amount, the Hausdorff dimension of galaxies is simply related through Eq. (2) to the correlation function of the primordial density field (if the galaxies are linearly biased tracers of mass). In the strongly non-linear regime, according to the model of self-similar gravitational clustering (e.g. Peebles 1993), the exponent of the correlation function and then the Hausdorff dimension is still determined by the power spectrum: $\gamma = 3(3+n)/(5+n)$. Alternatively, according to the thermodynamical model of self-gravitating systems (see Saslaw 1985), in the strongly non-linear regime the galaxy distribution loses memory of its initial conditions; in the case of thermodynamical equilibrium Saslaw has shown that $\gamma = 2$ is reached.

Gurzadyan & Kocharyan (1991) have approached this problem from the rather general position of the theory of dynamical systems: assuming the hypothesis that a gravitating system is subject to Kolmogorov instability, they derive the following formula:

$$d_H \simeq d(1 - ((d-1)/d)\exp(-T/\tau_{\rm GS})), \qquad (3)$$

where $d_H$ is the Hausdorff dimension, $d$ is the topological dimension (3 for usual space), $T$ is the Hubble time and $\tau_{\rm GS}$ is the Gurzadyan-Savvidy relaxation time (Gurzadyan & Savvidy 1984, 1986) of a system of gravitating objects:

$$\tau_{\rm GS} = \left(\frac{15}{4}\right)^{2/3} \frac{1}{2\pi\sqrt{2}} \frac{v}{Gmn^{2/3}}. \qquad (4)$$

Here $v$ is the velocity dispersion, $m$ the mass of the objects and $n$ is the number density; this last quantity, which is scale-dependent for a pure fractal, has to be understood for our purposes as related to a system with a well-defined size, e.g. a globular cluster, while the fractal behaviour refers to smaller scales. Eq. (3) gives us a very general dynamical explanation of the Hausdorff dimension of a galaxy system, and, furthermore, it connects the geometric observable $d_H$ to the dynamical properties of the system itself.

Gurzadyan & Kocharyan (1991) have also shown that the fractal dimension of the distribution of a galaxy system in 6D phase-space can be connected, by means of the Kaplan-Yorke hypothesis, to the so-called Lyapunov characteristic numbers; these, applied to the large-scale structure investigation, can be a precious tool for giving a precise and model-independent characterization of the matter distribution, and for obtaining

*email: monaco@tsmi19.sissa.it

Finally, this same formalism can be used to give a precise characterization of the results of numerical simulation.

## 2. Analysis

The actual validity of the Gurzadyan-Savvidy relaxation time has recently been supported observationally by Vesperini (1992) for globular clusters and by Pucacco (1992) for elliptical galaxies. The role of Gurzadyan-Savvidy relaxation for compact groups of galaxies is discussed in Mamon (1993). Our aim here is to check observationally the validity of Gurzadyan-Kocharyan relation (Eq. 3) for a small sample of well-studied rich clusters of galaxies.

Actually, clusters of galaxies are very ill-defined objects, whose dark mass component, largely dominant, is poorly known. Moreover, the properties of the galaxy members are poorly known also (mass-to-light ratio, luminosity function), so that parameters like the mass or the density of cluster galaxies are hard to estimate. Finally, quantities like the density of galaxies, which is scale-dependent for a pure fractal, are to be referred to the cluster size (e.g. the core radius), which is a somewhat uncertain quantity. Nonetheless, making reasonable assumptions it is possible to give an estimate of the parameters $\tau_{GS}$ and $d_H$.

We do not know if the dark mass distribution is clumpy or smooth. However, it is reasonable to assume that, at small scales, all the mass has collapsed in small clumps, whose minimum mass is determined by dissipative processes which took place in the early Universe (see, e.g., Efstathiou 1990). In the case of a cold or mixed (cold plus hot) dark matter universe, this minimal mass is reasonably of the order of the typical mass of globular clusters. Therefore one expects a galaxy cluster to be made up of clumps, luminous or dark, their masses ranging from about $10^6 M_\odot$ to about $10^{13} M_\odot$ according to some distribution function (the mass function). We assume, as a first approximation, that the mass function is independent of local density, which implies that light traces mass.

**Table 1.** The four clusters of galaxies

| Cluster | $cz$ (km/s) | R | Ap.(Mpc) | N |
|---|---|---|---|---|
| A 401 | 21900 | 2 | 1.0 | 94 |
| A 426 | 5180 | 2 | 1.5 | 91 |
| A 1656 | 7210 | 2 | 0.75 | 156 |
| A 1795 | 19130 | 2 | 1.2 | 78 |

To calculate the relaxation time $\tau_{GS}$ one has to estimate the velocity dispersion $v$, the mean density $n$ and the mean mass $m$ of the clumps in the cluster. If the cluster has undergone a violent relaxation (Lynden-Bell 1967), the clumps will have reached some degree of equipartition of velocities, apart from the most massive ones, which could have reached some degree of energy equipartition through dynamical friction; some evidence in this sense has been found by Biviano et al. (1992). As a consequence, one can say that the velocity dispersion of the brightest galaxies, which is the quantity one can measure, is a meaningful estimate of the true velocity dispersion $v$.

A more severe problem is the estimation of the number density of cluster clumps. The most reliable measure of the density of bright galaxies in clusters are the counts $N_{0.5}^c$ of galaxies brighter than $m_3 + 2$, where $m_3$ is the apparent magnitude of the third brightest galaxy in the cluster, within 0.5 $h^{-1}$ Mpc from the cluster center (Bahcall 1977, 1981). We assume that the actual density of clumps is proportional to the Bahcall counts: $n = \alpha N_{0.5}^c$; actually, this is equivalent to saying that the mass function is the same for all the clusters, which is a delicate but reasonable assumption, as we will use clusters of the same richness class.

We can sum up all our ignorance in the determination of the quantity $\tau_{GS}/T$ in the following way:

$$\tau_{GS}/T = 1.46 \, v \, (N_{0.5}^c)^{-2/3} p, \qquad (5)$$

($v$ in km/s), where $p$ is an "uncertainty" parameter whose value is

$$p = 1/(mh\alpha^{2/3}) \qquad (6)$$

($m$ in units of $10^{10} M_\odot$). The dimensionless parameter $\alpha$ will include also a correction of order unity for projection effects in the Bahcall counts. We note moreover that the Gurzadyan-Kocharian relation predicts $d_H = 1$ for $T \to 0$, while the galaxy field of a protocluster will have in general a different initial value of the Hausdorff dimension; however, the predicted tendency of an increase in $d_H$ for evolved systems should remain valid, thus, given the indeterminacy in estimating $p$, we will not take into account this problem.

The stability of $p$ for different clusters of the same richness class is the only assumption we need to continue. However, to obtain an order-of-magnitude estimate of $p$ we need to assume a form for the mass function. Following Giuricin et al. (1993), we assume that the total mass of a galaxy $M$ is given by

$$M \propto L^{0.5 \pm 0.1} \qquad (7)$$

(Ashman et al. 1993; Bertola et al. 1993); if the luminosity function is of the Schechter (1976) type, the mass function will be of the form:

$$n(M) \propto (M/M^*)^{-1.8 \pm 0.2} \exp(-(M/M^*)^2), \qquad (8)$$

with $M^* \sim 1.5 \times 10^{12} h^{-1} M_\odot$. We assume that the actual mass distribution of matter clumps is represented by Eq. (8), up to the cutoff mass of $10^6 M_\odot$. With this mass function one can calculate $\alpha$ as the fraction of objects with masses larger than the mass of a galaxy with magnitude $m_3$, $m$ as the average mass of the objects, and finally $p$ from Eq. (6), assuming $h = 1$ for simplicity. We have found that a reasonable order of magnitude for $p$ is $10^{-2}$; it depends weakly on the lower cutoff of the mass function (roughly as the cubic root of the mass cutoff), but strongly on the power-law exponent of the mass function: changing it from $-1.6$ to $-2.0$, $p$ changes by nearly two orders of magnitude!

Given this indeterminacy in $p$, it is convenient to determine the Hausdorff dimension of the cluster galaxies by calculating their two-point correlation function. According to Eq. (3), if the relaxation time is of the same order of the Hubble time, we expect $d_H$ to be greater than 1 (1.2 actually). In this case, from Eq. (3) we can get a value for $\tau_{GS}/T$ which, when compared with the same one obtained from Eq. (5), will give us value for the parameter $p$. We expect this to be the same for our clusters and of order $10^{-2}$. Of course one can not exclude the possibility that a modulation in the actual value of $p$ conspires to construct a picture consistent with the theory outlined above; further studies will be able to detect such conspirancies and/or to exclude them.

**Table 2.** Results

| Cluster | $N_{0.5}^c$ | $v$ (km/s) | $(\tau_{GS}/T)_{Eq.(5)}$ | $d_H$ | $(\tau_{GS}/T)_{Eq.(3)}$ | $p_{est} \times 10^2$ |
|---|---|---|---|---|---|---|
| A 401 | 34 | 1254 | $174 \times p$ | $1.43 \pm 0.10$ | $4.13^{+1.42}_{-0.88}$ | $2.4^{+1.6}_{-0.8}$ |
| A 426 | 32 | 1253 | $181 \times p$ | $1.57 \pm 0.09$ | $2.98^{+0.66}_{-0.48}$ | $1.6^{+0.9}_{-0.4}$ |
| A 1656 | 28 | 1049 | $166 \times p$ | $1.71 \pm 0.13$ | $2.28^{+0.64}_{-0.44}$ | $1.4^{+0.8}_{-0.5}$ |
| A 1795 | 27 | 899 | $146 \times p$ | $1.73 \pm 0.20$ | $2.20^{+1.14}_{-0.60}$ | $1.5^{+1.3}_{-0.6}$ |

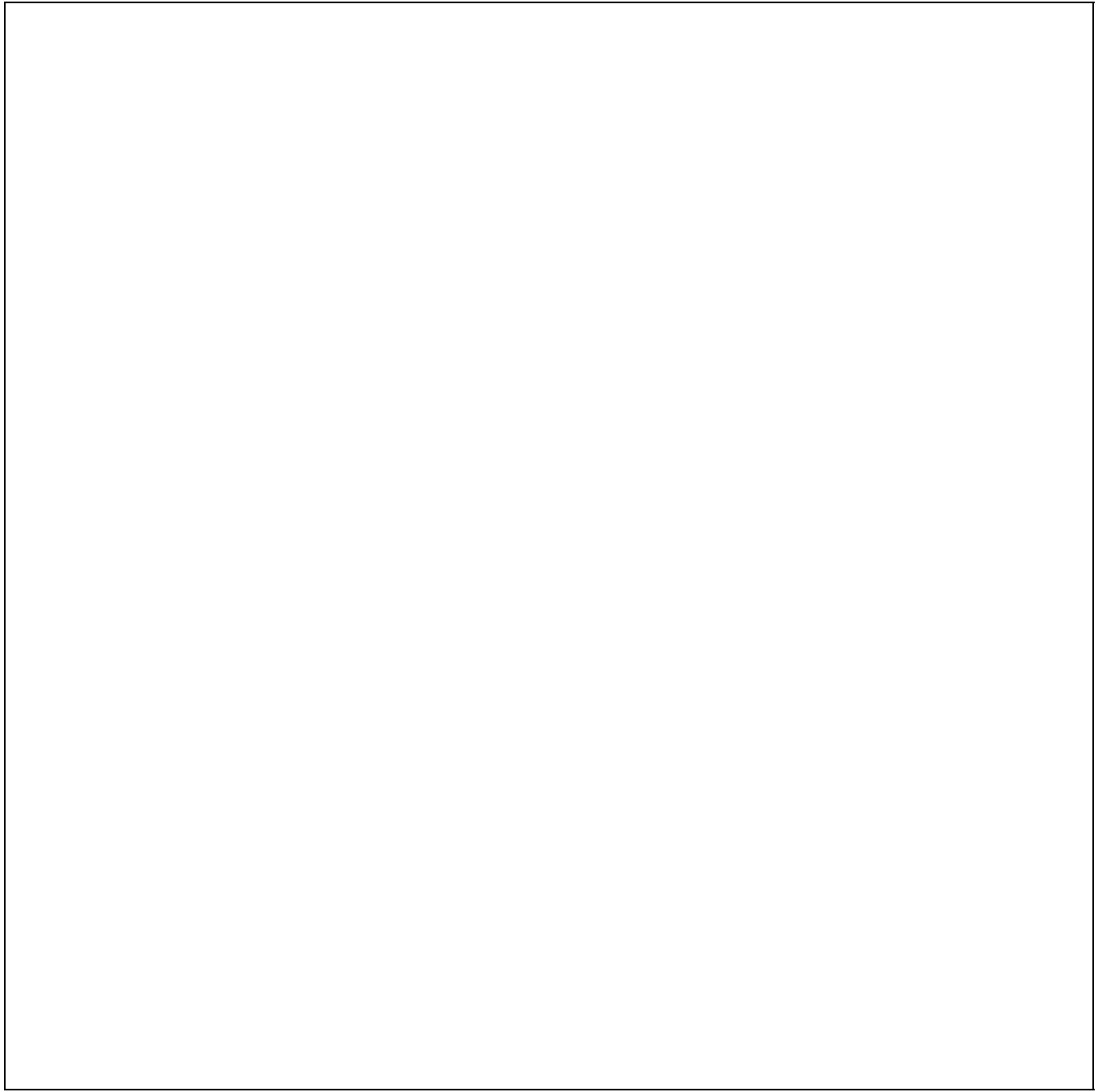

**Fig. 1.** The angular two-point correlation functions for the four clusters; the power-law fit of the small-scale part is shown

In order to calculate both $v$ and $d_H$ from the same data, we have taken samples of cluster galaxies with known redshifts for four rich clusters (Abell richness class R=2): Abell 401, Abell 426 (Perseus), Abell 1656 (Coma) and Abell 1795 (see Table 1). Data for Abell 401 and Abell 1795 were taken by Hill & Oegerle (1993); no information on magnitude completeness is available. Data for Abell 426 were taken by Kent & Sargent (1983), and are complete for $B_T < 14$. Data for Abell 1656 were taken by Kent & Gunn (1982), and are complete for $B_T < 14.5$. Robust estimates of the velocity dispersions were obtained using the methods developed in Girardi et al. (1993); the same methods allow the exclusion of the recognized interlopers. As mentioned previously, central densities $N_{0.5}^c$ have been taken by Bahcall (1981). We have considered the galaxies within physical apertures of 1.0, 1.5, 0.75 and 1.2 $h^{-1}$ Mpc respectively for the four clusters and verified that our results do not change significantly using different apertures. The lack of dependence of our results on the magnitude incompleteness has been checked for Abell 426 and Abell 1656 by verifying that the subsamples complete in magnitude give essentially the same results as the original ones. The same has not been done for Abell 401 and Abell 1795, because of the lack of magnitude information; however, we feel rather confident in saying that magnitude incompleteness ought not to be a severe problem in our analysis.

For each cluster we have calculated the angular correlation function $w(\theta)$ by means of the equation $w(\theta) = N_{\rm pp}/N_{\rm rp} - 1$, where $N_{\rm pp}$ is the number of pairs at angular distance $\theta$ and $N_{\rm rp}$ is the number of cross-pairs, again at distance $\theta$, between the sample and a given sample of an equal number of galaxies randomly distributed within the same aperture; the results have been averaged over 200 realizations of random samples. This cross-correlation procedures allows a minimization of border effects. Moreover, bootstrap errors have been obtained following Mo et al. (1992). We have verified that the results do not change when the binning of angular distances is varied. In Fig. 1 we show the obtained correlation functions. In every case we have noted a different behaviour at smaller and larger scales: at smaller scales, up to an angular scale of a few tenths of degree, the correlation function follows a power-law. At larger scales the correlation function steepens and then becomes negative, just due to the geometry of the cluster profile (and the integral constraint; see e.g., Peebles 1993). In the small-scale range defined above we have fitted the correlation function with a power-law of exponent $-\beta$ by means of a weighted $\chi^2$-minimization; in any case the $\chi^2$ probability of the fit resulted at least as good as 99%. Finally, using the relation $\gamma = 1 + \beta$ (e.g. Peebles 1993) and Eq. (2) we have estimated the Hausdorff dimension as $d_H = 2 - \beta$; Table 2 lists the results for the four clusters.

In all cases the Hausdorff dimensions are marginally or significantly greater than 1.2. This is consistent with Eq. (3), which predicts a larger Hausdorff dimension for systems whose dynamical time is just slightly larger than the Hubble time. Now we can verify whether the agreement is just qualitative or if Eqs. 3 and 4 give a consistent dynamical picture. Using Eq. (3) and the obtained values of $d_H$ we have calculated $\tau_{\rm GS}/T$; a comparison of these values with the ones obtained by means of Eq. (5) gives us an estimate $p_{\rm est}$ of the uncertainty parameter $p$; errors on $p_{\rm est}$ have been calculated allowing a 20% error on the dynamical estimate of $\tau_{\rm GS}/T$. As a matter of fact, these values of $p$ (see Table 2) are consistent among themselves within errors and are of the order of magnitude predicted. Hence, despite of the number of assumptions made and of the large errors in the measures, we can conclude that the Gurzadyan-Kocharyan relation gives a dynamical description of the space distribution of cluster galaxies which is consistent with our data. The study of large samples of galaxies in clusters and in other systems will be able to give more precise answers.

*Acknowledgements.* The author thanks V.G. Gurzadyan for enlightening comments, M. Plionis for his precious help in calculating the correlation functions, M. Cavaglià for useful comments, F. Mardirossian, A. Biviano, M. Girardi, G. Giuricin and M. Mezzetti for useful discussions on the observational data, and the referee, P. Coleman, for interesting suggestions.

# References

Ashman K.M., Persic P., Salucci M., 1993, MNRAS 260, 610
Bahcall N.A., 1977, ApJ 217, L77
Bahcall N.A., 1981, ApJ 247, 787
Bahcall N.A., 1988, ARA&A 26, 631
Bardeen J.M., Bond J.R., Kaiser N., Szalay A.S., 1986, ApJ 304, 15
Bertola F., Pizzella A., Persic M., Salucci P., 1993, ApJ 416, L45
Biviano A., Girardi M., Giuricin G., Mardirossian F., Mezzetti M., 1992, ApJ 369, 35
Coleman P.H., Pietronero L., 1992, Phys. Rep. 231, 312
Efstathiou G., 1990, Cosmological Perturbations. In: Peacock J.A., Heavens A.F., Davies A.T. (eds.) Physics of the Early Universe. Edinburgh Univ. Press, Edinburgh
Girardi M., Biviano A., Giuricin G., Mardirossian F., Mezzetti M., 1993, ApJ 404, 38
Giuricin G., Mardirossian F., Mezzetti M., Persic M., Salucci P., 1993, Preprint SISSA 132/93/A
Gurzadyan V.G., Kocharyan A.A., 1991, Europh. Lett. 15, 801
Gurzadyan V.G., Savvidy G.K., 1984, Dokl.Akad.Nauk.SSSR 277, 69
Gurzadyan V.G., Savvidy G.K., 1986, A&A 160, 203
Hill J.M., Oegerle W.R., 1993, AJ 106, 831
Kaiser N., 1984, ApJ 284, L9
Kent S.M., Gunn J.E., 1982, AJ 87, 945
Kent S.M., Sargent W.L.W., 1983, AJ 88, 697
Lynden-Bell D., 1967, MNRAS 136, 101
Mamon G.A., 1993, Dynamical Theory of Groups and Clusters of Galaxies. In: Combes F., Athanassoula E. (eds.) Gravitational Dynamics and the N-body Problem, in press
Mo H.J., Jing Y.P., Börner G., 1992, ApJ 392, 452
Peebles P.J.E., 1993, Principles of Physical Cosmology. Princeton Univ. Press, Princeton
Pucacco G., 1992, A&A 259, 473
Saslaw W.C., 1985, Gravitational Physics of Stellar and Galactic Systems. Cambridge Univ. Press, Cambridge
Schechter P.L., 1976, ApJ 203, 297
Vesperini E., 1992, A&A 266, 215

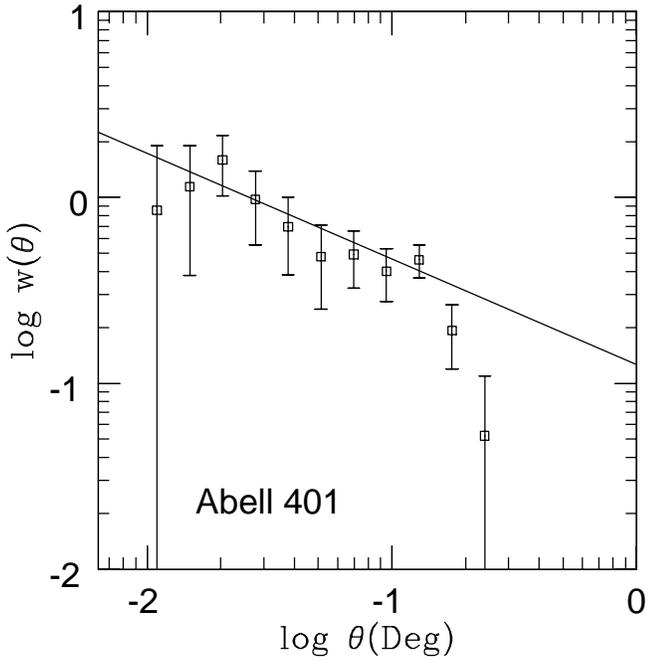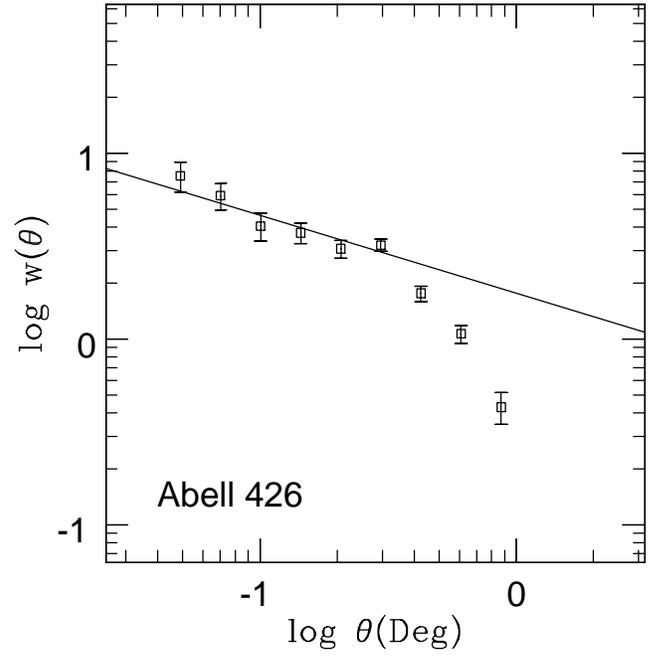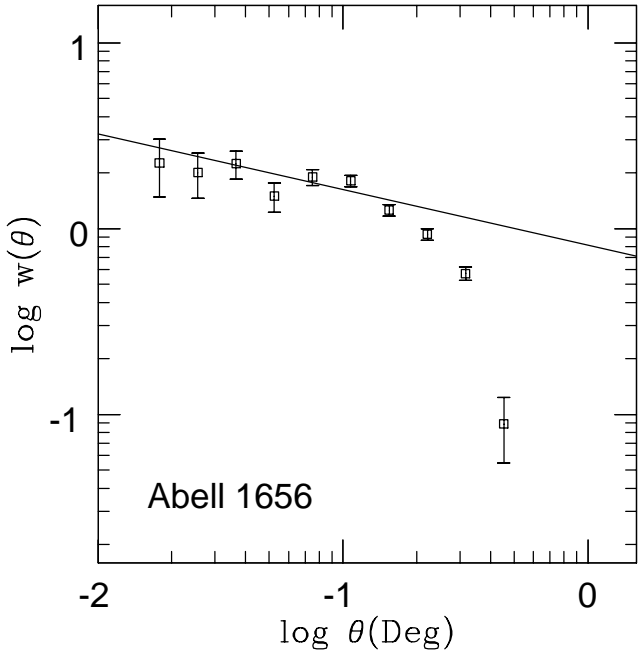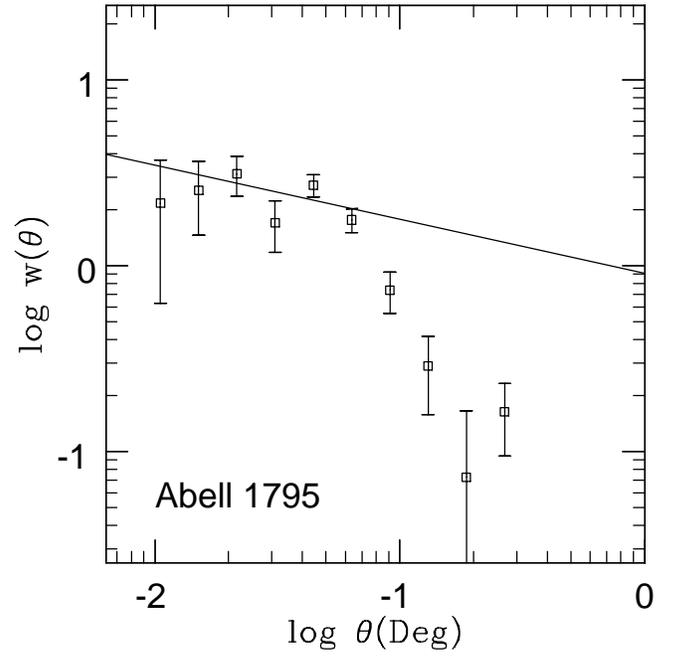